\documentclass[twocolumn,prl,showpacs,preprintnumbers,amsmath,amssymb,floatfix]{revtex4}

\usepackage{graphicx}
\usepackage{dcolumn}
\usepackage{bm}

\begin{document}

\preprint{YITP-06-45, RCNP-Th06025}

\title{Exotic hadrons in $s$-wave chiral dynamics}

%
%

\author{Tetsuo~Hyodo$^{a}$,
Daisuke~Jido$^{a}$, and
Atsushi~Hosaka$^b$}

\affiliation{$^a$Yukawa Institute for Theoretical Physics, 
Kyoto University, Kyoto 606--8502, Japan \\
$^b$Research Center for Nuclear Physics (RCNP),
Ibaraki, Osaka 567-0047, Japan.
}

\date{\today}
\begin{abstract}
    We study $s$-wave scattering of a hadron and a Nambu-Goldstone boson
    induced by the model-independent low energy interaction in the flavor
    SU(3) symmetric limit. Establishing the general structure of the low 
    energy interaction based on group theoretical arguments, we find that the
    interaction in the exotic channels are in most cases repulsive, and that 
    for possible attractive channels the interaction strengths are weak and 
    uniquely given independent of channel. Solving the scattering problem 
    with this interaction, we show that the attraction in the exotic channels
    is not strong enough to generate a bound state from the physically known 
    target hadrons. We also find that there are no attractive interaction in 
    the exotic channels in large $N_c$ limit.
\end{abstract}




\maketitle


Chiral symmetry is one of the fundamental symmetries in QCD with massless 
quarks and is spontaneously broken down to the diagonal flavor symmetry. The
low energy dynamics of the Nambu-Goldstone (NG) boson with other hadrons are
essentially governed by chiral symmetry and its spontaneous 
breaking~\cite{Weinberg:1979kz,WT}. Many recent studies of hadron resonances 
and exotic hadrons are based on the underlying QCD with the aid of chiral 
symmetry.

Conventionally, hadron resonances have been studied in the $S$-matrix theory.
There the scattering amplitude is obtained by considering the analyticity and
unitarity together with a partial wave decomposition valid in a limited range
of energies~\cite{Scatt}. One of the classical applications is the 
description of $\Lambda(1405)$~\cite{Lambda}, where $\Lambda(1405)$ emerges 
as a Feshbach resonance in the $s$-wave coupled channel problem in the 
$\pi\Sigma$ and $\bar K N$ channels. The recent development on this line is 
to determine the low energy scattering amplitude by chiral symmetry, 
reproducing various baryon resonances as dynamically generated 
states~\cite{KO,Oller:2000fj,Lutz:2001yb}. Since it was shown that the 
resonances obtained in the coupled channel approach became bound states of 
single channels in the flavor SU(3) limit~\cite{Octet,Kolomeitsev:2003kt,
Sarkar:2004jh,Lutz:2003jw,Kolomeitsev:2003ac}, the origin of the physical 
resonances may be studied by the bound states in the SU(3) limit.

The exotic hadrons have the flavor which cannot be reached by $\bar qq$ or 
$qqq$, and must have more valence quarks than the ordinary hadrons. The 
pentaquark $\Theta^{+}$~\cite{Nakano:2003qx} is one of the possible 
candidates. In spite of continuous experimental efforts to search for exotic 
hadrons, existence of such a state has not been clearly established. Almost 
complete absence of exotic states in hadron spectrum is, however, highly 
nontrivial. Theoretically, there has been no simple way to exclude the exotic
states in effective models describing the ordinary hadrons well, even more 
the fundamental theory of QCD does not forbid the exotic states. 

In this paper, we propose an attempt to approach the problem of exotic 
hadrons based on the symmetries of the underlying QCD, where we study 
$s$-wave scattering problem of the NG boson from a target hadron in the SU(3)
limit. Firstly, based on group theoretical arguments, we show that 
interactions in most of exotic channels are repulsive and that attractive 
interactions appear only in some limited number of channels with a universal
strength. After establishing the general features of the low energy 
interaction, we examine whether the attraction is strong enough to generate 
an exotic state as a bound state of the NG boson.  


Let us consider the interaction of the NG boson with a target hadron~($T$).  
According to the current algebra of the three flavor chiral symmetry, the 
scattering amplitude of the NG boson from a hadron in the low energy limit is
unambiguously expressed by the hadron matrix element of the conserved vector 
current of the flavor symmetry. Thus the invariant amplitude $\cal M$ can be 
written in a model-independent way as  
\begin{equation}
  {\cal M}_{\alpha} = \frac{1}{f^{2}}\frac{p\cdot q}{2 M_{T}} 
  \left\langle {\bm F}_{T} \cdot {\bm F}_{\rm Ad} \right\rangle_{\alpha}
  + {\cal O}\left((m/M_{T})^{2}\right) \label{eq:invME} ,
\end{equation}
with the decay constant $f$, the masses of the NG boson $m$ and of target 
hadron $M_{T}$, and the momenta $p$ and $q$ for the hadron and the NG boson.
${\bm F}_{T}$ and ${\bm F}_{\rm Ad}$ stand for the flavor SU(3) generators in
the representations of the hadron $T$ and the NG boson of the adjoint (Ad) 
representation, respectively. Finally, due to flavor SU(3) symmetry, the 
matrix element $\cal M$ is diagonal in the SU(3) representation $\alpha$ of 
the $s$ channel hadron-NG boson system; $\langle ... \rangle_\alpha \equiv 
\langle \alpha | ... |\alpha \rangle$. All the above is the essence of the 
celebrated Weinberg-Tomozawa theorem~\cite{WT}. 

For a baryon target, we obtain the $s$-wave scattering amplitude as
\begin{equation}
    V_{\alpha }
    =-\frac{ \omega}{2f^2}C_{\alpha,T}  , 
    \label{eq:WTint}
\end{equation}
where $C_{\alpha,T}\equiv - \left\langle 2{\bm F}_{T} \cdot {\bm F}_{\rm Ad} 
\right\rangle_{\alpha}$ and $\omega$ denotes the energy of the NG boson. In 
this equation, the spin averaged sum has been taken as $V\equiv \frac{1}{2} 
\sum_{\sigma} \bar u {\cal M} u$ with the spinor normalization 
$\bar u u =1$. This is equivalent to the leading term of the chiral 
perturbation theory and known as the Weinberg-Tomozawa (WT) term. It is found
that the scattering amplitude~\eqref{eq:WTint} can be applied also to the 
meson target in the heavy mass approximation~\cite{hyodoprep}. 

Apart from the common factor in Eq.~\eqref{eq:WTint}, the coupling strength 
$C_{\alpha,T}$ is determined only by specifying the representations of the 
system $\alpha$ and the target $T$:
\begin{equation}
    C_{\alpha,T}
    = - \left\langle 2{\bm F}_{T} \cdot {\bm F}_{\rm Ad}  
     \right\rangle_{\alpha} 
    =  C_2(T)-C_2(\alpha)+3 ,
    \label{eq:WTintfinal}
\end{equation}
where $C_2(R)$ is the quadratic Casimir of SU(3) for the representation $R$,
and we use $C_{2}({\rm Ad}) = 3$ for the adjoint representation of the NG 
boson. In our notation, a negative $C_{\alpha,T}$ leads to a repulsive 
interaction, whereas a positive $C_{\alpha,T}$ gives an attractive 
interaction.   

Let the target hadron belong to an arbitrary SU(3) representation $[p,q]$ in 
the tensor notation. Possible representations $\alpha$ for the scattering 
channels are obtained in the irreducible decomposition of direct product of 
$[p,q]$ and the adjoint representation $[1,1]$ of the NG~boson:
\begin{eqnarray}
  \lefteqn{
    [p,q]\otimes [1,1] =    [p+1,q+1]   
    } && \nonumber \\   && 
    \oplus [p+2,q-1]
    \oplus [p-1,q+2]
    \oplus [p,q]
    \oplus [p,q] \nonumber \\ &&
    \oplus [p+1,q-2]
    \oplus [p-2,q+1] 
     \oplus [p-1,q-1] ,
    \label{eq:generalpq}
\end{eqnarray}
where the labels of representations $[a,b]$ satisfies $a, b \geq 0$, and one 
of the two $[p,q]$ representations on the right hand side satisfies $p\geq 1$
and the other $q\geq 1$. 

We evaluate the coupling strength $C_{\alpha,T}$ for the channel~$\alpha$ 
using Eq.~\eqref{eq:WTintfinal} and the quadratic Casimir $C_2([p,q])=
(p^2+q^2+pq+3(p+q))/3$. We show a general expression for the coupling 
strength of the WT interaction for arbitrary representations in the second 
column of Table~\ref{tbl:ECtable}. Since $p$ and $q$ are nonnegative integer,
$C_{\alpha,T}$ takes an integer value whose sign is determined for a given 
$\alpha$ except for $[p+2,q-1]$ and $[p-1,q+2]$. 

\begin{table}[tbp]
    \centering
    \caption{Properties of the WT interaction in the channel~$\alpha$ of the 
    NG boson scattering on the target hadron with the $T=[p,q]$ 
    representation. The coupling strengths of the WT term is denoted as 
    $C_{\alpha,T}$, $\Delta E$ is the differences of the exoticness $E$ 
    between the channel $\alpha$ and the target hadron $T$, and 
    $C_{\alpha,T}(N_c)$ denotes the coupling strengths for arbitrary $N_c$.}
    \begin{ruledtabular}
    \begin{tabular}{cccc}
	$\alpha$  
	& $C_{\alpha,T}$ 
	& $\Delta E$ 
	& $C_{\alpha,T}(N_c)$  \\
        \hline
        $[p+1,q+1]$ 
	& $-p-q$ 
	& 1 or 0 
	& $\frac{3-N_c}{2}-p-q$  \\
        $[p+2,q-1]$ 
	& $1-p$ 
	& 1 or 0
	& $1-p$  \\
        $[p-1,q+2]$ 
	& $1-q$ 
	& 1 or 0 
	& $\frac{5-N_c}{2}-q$ \\
        $[p,q]$ 
	& $3$  
	& 0 
	& 3 \\
        $[p+1,q-2]$ 
	& $3+q$ 
	& 0 or $-1$ 
	& $\frac{3+N_c}{2}+q$  \\
        $[p-2,q+1]$ 
	& $3+p$  
	& 0 or $-1$ 
	& $3+p$ \\
        $[p-1,q-1]$ 
	& $4+p+q$  
	& 0 or $-1$ 
	& $\frac{5+N_c}{2}+p+q$ \\
    \end{tabular}
    \end{ruledtabular}
    \label{tbl:ECtable}
\end{table}

We are interested in exotic channels in the scattering of the NG boson and a
target hadron. To specify the exotic channels, it is convenient to define the
exoticness $E$ of $[p,q]$ as in Refs.~\cite{Exotic,hyodoprep} in terms of the
number of valence quark-antiquark pairs to compose the given flavor multiplet
$[p,q]$ and the baryon number $B$ carried by the $u$, $d$, and $s$ quarks 
(the number of the heavy quarks is not included in this definition of the 
baryon number). We consider the light baryons ($qqq$), heavy baryons ($qqQ$),
and heavy mesons ($q\bar{Q}$) as the target hadrons. For $B>0$ the exoticness
$E$ is given by~\cite{hyodoprep}
\begin{equation}
    E=\epsilon\theta(\epsilon)+\nu\theta(\nu) ,
    \label{eq:exoticness}
\end{equation}
with the step function $\theta(x)$ and the quantities $\epsilon$ and $\nu$
defined by
\begin{equation}
    \epsilon \equiv
    \dfrac{p+2q}{3}-B, \quad 
    \nu\equiv\dfrac{p-q}{3}-B .
    \nonumber
\end{equation}
The detailed derivations will be given in Ref.~\cite{hyodoprep}.

The {\rm more} exotic channel is characterized by $\Delta E = 1$ where 
$\Delta E$ denotes the difference between the exoticness parameters $E$ of 
the channel $\alpha$ and that of the target hadron $T$. $\Delta E$ can take 
the values $+1$, $0$, and $-1$ when the quark-antiquark pair $q\bar{q}$ of 
the NG boson is added to the target hadron. The possible values of $\Delta E$
for all channels are given in the third column of Table~\ref{tbl:ECtable}. We
find that $\Delta E=1$ is achieved when one of the following conditions is 
satisfied:
\begin{enumerate}
    \item[(i)]  $\Delta \epsilon=1$, $\Delta \nu = 0$, 
    $\epsilon_T\geq 0$,
    \item[(ii)]  $\Delta \epsilon=0$, $\Delta \nu = 1$,
    $\nu_{T}\geq 0$,
    \item[(iii)]  $\Delta \epsilon=1$, $\Delta \nu = -1$,
    $\nu_{T}\leq 0$,
\end{enumerate}
with $\Delta \epsilon =\epsilon_{\alpha}- \epsilon_{T}$ and $\Delta 
\nu=\nu_{\alpha}- \nu_{T}$. The case (i) is satisfied for $\alpha=[p+1,q+1]$ 
but the interaction is always repulsive for this channel. The case (ii) is 
satisfied for $\alpha=[p+2,q-1]$ and $\nu_{T}\geq 0$, but the attraction is 
realized with $p=0$. This leads to a negative baryon number $B\leq -q/3$ 
because of $\nu_{T}\geq 0$. The case (iii) is satisfied for 
$\alpha=[p-1,q+2]$, where the interaction can be attractive only when $q=0$.
In this case, the strength is always $C_{\alpha,T}=1$ and the condition 
$\nu_T\leq 0$ gives $p\geq 3B$.

Thus, to summarize, we find that most of exotic channels are repulsive and 
the attractive interaction is realized only in the channel $\alpha=[p-1,2]$ 
of the NG boson scattering on the target hadron $T=[p,0]$ $(p\geq 3B)$ with 
the universal strength
\begin{equation}
    C_{\text{exotic}}=1  .  \label{eq:Exoticattraction}
\end{equation}
It is interesting that the attractive interaction for the exotic channel 
takes place with the smallest strength of the WT term, only when the target 
hadron belongs to the symmetric representation $[p,0]$. 

The next question is whether the attractive interaction 
Eq.~\eqref{eq:Exoticattraction} is strong enough to provide an exotic state.
For this purpose, we solve scattering problem with the WT interaction treated
beyond the perturbation. The scattering equation is solved as a single 
channel problem for each $\alpha$ in the SU(3) limit. There, enough 
attraction generates a bound state with the flavor quantum number~$\alpha$.

In order to obtain the relevant scattering amplitude we impose the elastic 
unitarity condition in the N/D method. This procedure has been shown to be 
equivalent to solving the scattering equation with the WT amplitude as the 
kernel interaction~\cite{Oller:2000fj}. Then we obtain the scattering 
amplitude of the NG boson and the target hadron in the channel $\alpha$ as
\begin{equation}
    t_{\alpha}(\sqrt{s})=
    \frac{1}{1-V_{\alpha}(\sqrt{s})G(\sqrt{s})}V_{\alpha}(\sqrt{s}) ,
    \label{eq:ChUamp}
\end{equation}
as a function of the center-of-mass energy $\sqrt{s}$. Here $V_{\alpha}
(\sqrt{s})$ denotes the WT interaction \eqref{eq:WTint}, and $G(\sqrt{s})$ is
given by the once-subtracted dispersion integral
\begin{equation}
    G(\sqrt{s})
    =-\tilde{a}(s_0)
    -\frac{1}{2\pi}
    \int_{s^{+}}^{\infty}ds^{\prime}
    \left(
    \frac{\rho(s^{\prime})}{s^{\prime}-s}
    -\frac{\rho(s^{\prime})}{s^{\prime}-s_0}
    \right) ,
    \label{eq:loop_s}
\end{equation}  
with the phase space integrand $\rho(s)=2M_{T}\sqrt{(s-s^+)(s-s^-)}/(8\pi s)$
and $s^{\pm}=(m\pm M_{T})^2$. The subtraction constant $\tilde{a}(s_0)$ is 
not determined within the N/D method. 

To fix $\tilde{a}(s_0)$, we take the prescription given in 
Refs.~\cite{Igi:1998gn,Lutz:2001yb}:
\begin{equation}
    G(M_{T})=0  ,
    \label{eq:regucond} 
\end{equation}
which is equivalent to $t_{\alpha}(\sqrt s)=V_{\alpha}(\sqrt s)$ at 
$\sqrt s =M_{T}$. Since the WT term $V_{\alpha}$ has the crossing symmetry, 
this prescription restores the crossing symmetry approximately in the 
scattering amplitude~\eqref{eq:ChUamp} at energies around $\sqrt s =M_{T}$.
This is enough for our purpose, since the bound state energy $M_b$ is
calculated within  $M_{T}< M_b < M_{T}+m$. The 
prescription~\eqref{eq:regucond} is realized in Eq.~\eqref{eq:loop_s} by 
choosing $\tilde{a}(s_{0})=0$ at $s_0=M_{T}^2$.

The bound state is expressed as a pole of the scattering 
amplitude~\eqref{eq:ChUamp}. To find the bound state, we define a function 
$D$ as the denominator of the amplitude:
\begin{equation}
    D(\sqrt{s})
    \equiv 1-V_{\alpha}(\sqrt{s})G(\sqrt{s}) .
    \label{eq:Ddef}
\end{equation}
The bound state energy $M_b$ should be within the energies between the target
mass and the scattering threshold. Thus if bound states exist, the bound 
state energies are obtained by 
\begin{equation}
    D(M_b)=0 
    \quad \text{with} \quad  M_{T}< M_b < M_{T}+m .
    \label{eq:boundsolution}
\end{equation}

Now we can show that there is at most one solution to the above equation as
follows. The renormalization condition~\eqref{eq:regucond} requires 
$D(M_{T})=1>0$. Observing that both $V_{\alpha}(\sqrt s)$ and $G(\sqrt s)$ 
shown in Eqs.~\eqref{eq:WTint} and \eqref{eq:loop_s} are decreasing functions
of $\sqrt{s}$ for attractive interaction $C_{\alpha,T}>0$ in the energy 
region $M_{T}< \sqrt{s} <M_{T}+m$, we find that $D=1-V_{\alpha}G$ is also 
monotonically decreasing in this region. This means that, if $D(M_{T}+m)<0$ 
is satisfied, we have a unique solution of the 
equation~\eqref{eq:boundsolution}.

\begin{figure}[tbp]
\centerline{\includegraphics[width=7cm]{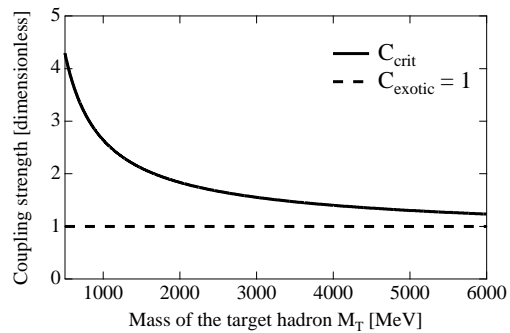}}
\caption{Critical coupling strength $C_{\text{crit}}$ for the averaged mass 
over the pseudoscalar octet mesons, $m=368$ MeV (Solid lines). The dashed 
line denotes the universal attractive coupling strength in exotic channels 
$C_{\rm exotic}=1$.}
\label{fig:critical}
\end{figure}

Thus, the smallest attractive coupling strength $C_{\rm crit}$ to produce a 
bound state of the NG boson in the target hadron is calculated so as to 
satisfy $D(M_{T}+m)=0$:
\begin{align}
    C_{\text{crit}}= \frac{2f^2 }{m\bigl(-G(M_{T}+m)\bigr)} .
    \label{eq:WTcritical}
\end{align}
If the coupling strength is smaller than $C_{\text{crit}}$ for a given NG 
boson mass $m$ and the target mass $M_{T}$, there is no bound state. Shown in
Fig.~\ref{fig:critical} is $C_{\text{crit}}$ as functions of $M_{T}$ with 
$f=93$ MeV. We use the average mass of the octet mesons ($\pi$, $K$, $\eta$)
$m=368$, since the observed pseudoscalar meson masses can be reproduced well 
by the averaged mass with the linear order of SU(3) breaking. We also plot 
$C_{\rm exotic}=1$, which is the universal strength of the attraction in 
exotic channels.

As seen in  Fig.~\ref{fig:critical}, it is clear that $C_{\rm exotic}=1$ is 
not enough to bind the NG boson in the target hadron with the mass $M_{T}<6$ 
GeV, where possible target hadrons have been observed in experiments. The 
critical coupling strength $C_{\text{crit}}$ is monotonically decreasing as 
the target mass $M_{T}$ increases. Therefore, it will become smaller than 1 
at sufficiently large $M_{T}$. Quantitatively, to have a bound state for the 
exotic channel $C_{\rm exotic}=1$, the mass of the target hadron $M_{T}$ 
should be larger than 14 GeV for $m=368$ MeV and $f=93$ MeV. Exotic states 
can appear as the bound states if a stable heavy hadron exists in this energy
region. 

If we use a heavier NG boson mass, the critical coupling $C_{\rm crit}$ is 
smaller than the present case and the NG boson could be bound in the exotic 
channel. For instance, with $m=$ 500 MeV an exotic bound state appears with 
$M_{T} \simeq$ 2500 MeV as shown in Ref.~\cite{Lutz:2003jw}. 


In the present approach, exotic states are treated as quasi-bound states of 
a NG boson and a hadron on the same footing as the nonexotic resonances.
We find bound states in some nonexotic channels with enough strength of 
attractions in the SU(3) limit as listed in Table~\ref{tbl:binding}. The 
target mass is taken as the averaged mass of the observed hadrons in the 
multiplets. We also show in Table~\ref{tbl:binding} examples of resonances 
found in those channels, referring to the literature in which realistic 
calculations are performed with the SU(3) breaking. Considering the fact that
several known resonances have been properly generated by the chiral unitary 
approaches, our conclusion on the exotic states should be of great relevance.

\begin{table}[tbp]
    \centering
    \caption{Masses of the dynamically generated states in the SU(3) limit.
    The values of the masses $M_T$ and $M_b$ are given in units of MeV. The 
    SU(3) representations are written in their dimensions. Examples of the
    corresponding resonances with SU(3) breaking are shown in the last 
    column.}
    \begin{ruledtabular}
    \begin{tabular}{ccccccl}
        target hadron & $T$ & $M_T$ & $\alpha$ & $C_{\alpha,T}$
	& $M_b$& Examples \\
        \hline
        light baryon & $\bm{8}$ & 1151 & $\bm{1}$ & $6$ & 1450 &
	$\Lambda(1405)$~\cite{Octet}  \\
        && & $\bm{8}$ & $3$ & 1513 & $\Lambda(1670)$~\cite{Octet} \\
	& $\bm{10}$ & 1382 & $\bm{8}$ & $6$ & 1668
	& $\Lambda(1520)$~\cite{Sarkar:2004jh} \\
        && & $\bm{10}$ & $3$ & 1737 & $\Xi(1820)$~\cite{Sarkar:2004jh} \\
        charmed & $\overline{\bm{3}}$
	& 2408 & $\overline{\bm{3}}$ & $3$ & 2736
	& $\Lambda_c(2880)$~\cite{Lutz:2003jw} \\
	baryon & $\bm{6}$ & 2534 & $\overline{\bm{3}}$ & $5$ & 2804
	& $\Lambda_c(2593)$~\cite{Lutz:2003jw} \\
        && & $\bm{6}$ & $3$ & 2860 & - \\
	$D$ meson & $\bm{3}_c$ & 1900 & $\bm{3}$ & $3$ & 2240 &
	$D_s(2137)$~\cite{Kolomeitsev:2003ac} \\
	$B$ meson & $\bm{3}_b$ & 5309 & $\bm{3}$ & $3$ & 5600 & - \\
    \end{tabular}
    \end{ruledtabular}
    \label{tbl:binding}
\end{table}


Our discussion on the coupling strength of the WT term can be extended to 
baryons with the arbitrary number of color $N_{c}$. It is known that the WT 
term scales as $1/N_c$ in the large $N_c$ limit, since it contains $1/f^2$ 
and $f \propto \sqrt{N_c}$. This is the case that the coupling strength 
$C_{\alpha,T}$ has no $N_c$ dependence. Here we show that $C_{\alpha,T}$ 
does have the $N_c$ dependence in the case of the baryon target. Also in 
Ref.~\cite{Garcia-Recio:2006wb}, nontrivial $N_c$ dependence was reported for
the spin-flavor SU(6) extended WT term. An interesting discussion was 
recently made on interplay of the chiral and $1/N_c$ 
expansions~\cite{Cohen:2006up}. For arbitrary $N_c$, a baryon is constructed 
by $N_c$ quarks. The flavor multiplet of baryon $[p,q]$ for $N_{c}=3$ is 
conventionally extended to $[p,q+(N_c-3)/2]$ for arbitrary $N_c$ so as to fix
the baryon spin~\cite{Dashen:1993jt}, while the SU(3) representation of the 
meson does not change. Using Eq.~\eqref{eq:WTintfinal}, we obtain the 
coupling strength for arbitrary $N_c$, which are summarized in the last 
column of Table~\ref{tbl:ECtable}. We find that the possible attraction in 
exotic channels~\eqref{eq:Exoticattraction} for $N_{c}=3$ changes into 
repulsive in the large $N_c$ limit. This means that there is no way to 
provide bound states in exotic channels in the large $N_c$ limit, 
independently of the masses of the $NG$ boson and target baryons.


To conclude, we have investigated possibility to have exotic states in chiral
dynamics, focusing on the study of an $s$-wave bound state of a 
Nambu-Goldstone boson from a target hadron in the SU(3) limit. The chiral 
dynamics supplies the model independent coupling strength for the $s$-wave 
scattering of the NG boson and the target hadron at low energies. Based on 
the group theoretical arguments, we have shown that most of exotic channels 
are repulsive and possible attractive channels are weak with a universal 
strength $C_{\rm exotic}=1$. Exploiting the scattering theory with the WT 
term as the kernel interaction, we have proved that the strength of
attraction in the exotic channel of the WT term is not enough to generate 
bound states of the NG boson with the experimentally observed hadrons. We 
have also found that there are no attractive exotic channels in the large
$N_c$ limit. It should be worth noting that the present analysis does not 
exclude the existence of the exotic states formed by other mechanisms, for 
instance the genuine quark states. 

The authors are grateful to Prof.\ M.\ Oka for helpful discussion. T.H. 
thanks the Japan Society for the Promotion of Science (JSPS) for financial 
support. This work is supported in part by the Grant for Scientific Research 
(Nos.\ 17959600, 18042001, 16540252) and by Grant-in-Aid for the 21st Century
COE "Center for Diversity and Universality in Physics" from the Ministry of 
Education, Culture, Sports, Science and Technology (MEXT) of Japan.

\end{document}